\documentclass[aps,prd,twocolumn,superscriptaddress, showpacs,amsmath,amssymb,nofootinbib, preprintnumbers]{revtex4-1}
\usepackage{graphicx}

\usepackage{epstopdf}

\usepackage{color}

\newcommand{\red}{\textcolor{red}}

\newcommand{\be}{\begin{equation}}
\newcommand{\ee}{\end{equation}}
\newcommand{\ba}{\begin{eqnarray}}
\newcommand{\ea}{\end{eqnarray}}

\newcommand{\beq}{\begin{equation}}
\newcommand{\eeq}{\end{equation}}
\newcommand{\beqa}{\begin{eqnarray}}
\newcommand{\eeqa}{\end{eqnarray}}

\begin{document}
\title{Van der Waals black hole}

\author{Aruna Rajagopal}
\email{arajagopal@perimeterinstitute.ca}
\affiliation{Perimeter Institute, 31 Caroline St. N., Waterloo,
Ontario, N2L 2Y5, Canada}

\author{David Kubiz\v n\'ak}
\email{dkubiznak@perimeterinstitute.ca}
\affiliation{Perimeter Institute, 31 Caroline St. N., Waterloo,
Ontario, N2L 2Y5, Canada}
\affiliation{Department of Physics and Astronomy, University of Waterloo,
Waterloo, Ontario, Canada, N2L 3G1}

\author{Robert B. Mann}
\email{rbmann@sciborg.uwaterloo.ca}
\affiliation{Department of Physics and Astronomy, University of Waterloo,
Waterloo, Ontario, Canada, N2L 3G1}

\date{November 28, 2014}  

\begin{abstract}
In the context of extended phase space, where the negative cosmological constant is treated as a thermodynamic pressure in the first law of black hole thermodynamics, we find an asymptotically AdS metric whose thermodynamics matches  exactly that of the Van der Waals fluid.  However, we show that as a solution of Einstein's equations, the corresponding stress energy tensor does not obey any of the energy conditions everywhere outside of the horizon.  
\end{abstract}


\maketitle

\section{Introduction}
Due to the AdS/CFT correspondence, there has been a revival of interest in the physics of asymptotically AdS black holes in recent years; the main focus being that of understanding strongly coupled thermal field theories living on the AdS boundary.   
Even from a bulk perspective such black holes are quite interesting, their thermodynamics exhibiting  various phase transitions. A primary example is the thermal radiation/black hole first-order phase transition observed for Schwarzschild-AdS black hole spacetimes \cite{HawkingPage:1983}. 
Interestingly, when a charge and/or rotation are  added,  behaviour qualitatively similar 
to a Van der Waals fluid emerges \cite{ChamblinEtal:1999a,Cvetic:1999ne,CaldarelliEtal:2000,Niu:2011tb}. This analogy becomes 
even more complete in the extended phase space \cite{KubiznakMann:2012,Gunasekaran:2012dq}, where the cosmological constant is treated as a thermodynamic pressure $P$, 
\be
P=-\frac{\Lambda}{8\pi}=\frac{3}{8\pi l^2}\,,
\ee
 and is allowed to vary in the first law of black hole thermodynamics, 
\be\label{first}
\delta M=T \delta S+V \delta P+\dots,
\ee
while a quantity thermodynamically conjugate to $P$ is interpreted as a black hole thermodynamic volume \cite{KastorEtal:2009, Dolan:2010}
\be\label{V}
V=\bigg(\frac{\partial M}{\partial P}\bigg)_{S,\dots}\,.
\ee
 This allows one to write down a `black hole equation of state' $P=P(V,T)$ and compare it to the corresponding fluid equation of state, while
we identify the black hole and fluid temperatures $T\sim T_f$,  the black hole and fluid volumes $V\sim V_f$, and the cosmological and fluid pressures $P\sim P_f$. 

The Van der Waals fluid is described by the Van der Waals equation, which is a closed form two parameter equation of state:
\be\label{VdW}
T=\Bigl(P+\frac{a}{v^2}\Bigr)(v-b)\,, 
\ee
where $v$ denotes the specific volume of the fluid, $v=V/N$ with $N$ counting  fluid's degrees of freedom. The parameter $a>0$ measures the attraction in between the molecules of the fluid, and the parameter $b$ measures their volume.
The thermodynamics of charged and/or rotating AdS black holes have been shown to qualitatively mimic the behavior of this equation \cite{KubiznakMann:2012, Gunasekaran:2012dq}, including the existence of a small/large black hole first-order phase transition corresponding to a liquid/gas phase transition that eventually terminates at a critical point characterized by the standard mean field theory critical exponents. In both cases, the corresponding thermodynamic potential, the Gibbs free energy, 
displays the swallowtail catastrophe. 

Although remarkable, the analogy between the thermodynamics of charged/rotating AdS black holes and that of the Van der Waals fluid is only {\em qualitative}---the corresponding equations of state   
are not identical and prevent one from identifying the black hole parameters such as charge $Q$ or rotation $J$  with the fluid parameters 
$a$ and $b$. Unfortunately, this remains true for all other more complicated examples of black holes, possibly in higher or lower dimensions, which were found to demonstrate a qualitative Van der Waals behavior; for a recent review see \cite{Altamirano:2014tva}. 

 In this letter we turn the logic around and construct an asymptotically AdS black hole whose thermodynamics matches exactly that of the Van der Waals fluid. We show that, as a solution of Einstein's equations, the corresponding stress energy tensor does not obey any of the three standard energy conditions everywhere outside the horizon, though for certain values of
 the parameters it is possible to satisfy the energy conditions in a region near the horizon. 
  Such black holes are then likely unphysical.

\section{Constructing the solution}

In what follows we want to construct an asymptotically AdS black hole metric whose thermodynamics coincide exactly with the given fluid equation of state. For simplicity here we assume the  
static spherically symmetric ansatz
\ba\label{metric}
ds^2&=&-f dt^2+\frac{dr^2}{f}+r^2d\Omega^2\,, \\
f&=&\frac{r^2}{l^2}-\frac{2M}{r}-h(r,P)\,,  \nonumber
\ea
where the function $h$ is to be determined.

We further assume that such a metric is a solution of the Einstein field equations with a given energy-momentum source, $G_{ab}+\Lambda g_{ab}=8\pi T_{ab}$. 
In order the energy-momentum source be physically plausible, one should require it to satisfy certain conditions such as positivity of energy density and dominance of the energy density over pressure, known as energy conditions, e.g. \cite{poisson2004relativist}. Namely, the minimal requirement  one should impose   is the {\em weak energy condition}, demanding $T_{ab}\xi^a\xi^b\geq 0$ for any future-directed timelike vector $\xi$. Writing the stress energy tensor in an orthonormal basis, $T^{ab}=\rho e_0^a e^b_0+\sum_i p_i e^a_i e_i^b$, where $\rho$ stands for the energy density and $p_i$ denote principal pressures, the energy conditions imply ($i=1,2,3$)
\beqa\label{weak}
\mbox{Weak:}\quad &\rho\geq 0\,,\quad \rho+p_i\geq 0\,,\\
\label{strong}
\mbox{Strong:}\quad &\rho+\sum_i p_i\geq 0\,,\quad \rho+p_i\geq 0 \,,\\
\label{dominant}
\mbox{Dominant:}\quad & \rho \geq | p_i |\,.
\eeqa
In particular, for a metric ansatz we find (with prime denoting the derivative w.r.t. $r$)
\ba
\rho&=&-p_1=\frac{1-f-r f'}{8\pi r^2}+P\,,\nonumber\\
p_2&=&p_3=\frac{rf''+2f'}{16\pi r}-P\,.
\ea 
Once we determine $f$, we shall check the corresponding energy conditions above.

The ansatz \eqref{metric} implies that the ``mass'' of the black hole $M$ is related to the horizon radius $r_+$ according to 
\be
M=\frac{4}{3}\pi r_+^3P-\frac{h(r_+,P)r_+}{2}\,.
\ee
Imposing further the first law, \eqref{first}, the thermodynamic volume $V$ is determined from \eqref{V}.
Since the horizon area is given by $A=4\pi r_+^2$, we now define the  black hole `specific volume' as 
\be
v=k\frac{V}{N}\,,\quad N=\frac{A}{L_{pl}^2}\,,
\ee
where in accord with previous papers we set the constant (in $d=4$ dimensions) to be $k=\frac{4(d-1)}{d-2}=6$, 
and interpret $N$ as the `number of degrees of freedom' associated with the black hole horizon, with $L_{pl}$ being the Planck length
\cite{KubiznakMann:2012,Gunasekaran:2012dq, Altamirano:2014tva},
giving
\be\label{v}
v=\frac{k}{4\pi r_+^2}\Bigl[\frac{4}{3}\pi r_+^3-\frac{r_+}{2}\frac{\partial h(r_+,P)}{\partial P}\Bigr]\,.
\ee
Since we are in Einstein gravity, the entropy and the horizon area are related as $S=A/4$.
We also know that the black hole temperature reads 
\be\label{T}
T=\frac{f'(r_+)}{4\pi}=2r_+P-\frac{h(r+,P)}{4\pi r_+}-\frac{1}{4\pi}\frac{\partial h(r_+,P)}{\partial r_+}\,.
\ee
This can now be compared to any desired fluid equation of state, $T=T(v,P)$.

\section{Van der Waals black hole}
The discussion in the previous section can be applied to any desired equation of state. Let us now specify to the Van der Waals case \eqref{VdW}. That is, we compare the expression for $T$ \eqref{T} with \eqref{VdW}, to get
\be
2r_+P-\frac{h}{4\pi r_+}-\frac{h'}{4\pi}-\Bigl(P+\frac{a}{v^2}\Bigr)(v-b)=0\,,
\ee
where we substitute for $v$ the expression \eqref{v}. This represents a partial differential equation for $h(r,P)$ which 
gives a solution to our problem. 
 
In particular, we find a solution of this equation by employing the following ansatz:
\be\label{ansatzh}
h(r,P)=A(r)-P B(r)\,.
\ee
With this ansatz the above PDE becomes of the form $F_1(r) P+F_2(r)=0$, where $F_1$ and $F_2$ depend on functions $A$ and $B$ and their derivatives. Since both these parts have to vanish separately, we get a system of two ODEs for unknown functions $A$ and $B$. Solving first $F_1=0$ we find that (setting $k=6$) 
\be
B=\Bigl(C_1-\frac{8\pi}{3}\Bigr)r^2+4\pi b r\,.
\ee
Setting now the integration constant $C_1=\frac{8\pi}{3}$ [so that we preserve the AdS structure postulated in \eqref{metric}], we find that $F_2=0$ gives a solution
\be
A=-2\pi a+\frac{3\pi a b^2}{r(2r+3b)}+\frac{4\pi a b}{r}\log\Bigl(\frac{2r}{r_0}+3\frac{b}{r_0}\Bigr)\,, 
\ee
where $r_0$ is an integration constant with dimensions of length.  For simplicity setting $r_0=2b$ yields
\ba
f&=&2\pi a-\frac{2M}{r}+\frac{r^2}{l^2}\bigg(1+\frac{3}{2}\frac{b}{r}\bigg)-\frac{3\pi a b^2}{r(2r+3b)}\nonumber\\
&&-\frac{4\pi ab}{r}\log\bigg(\frac{r}{b}+\frac{3}{2}\bigg) \label{metf}
\ea
for the `Van der Waals black hole metric'. For $b<<r$ this expands as  
\ba
f&=&2\pi a-\frac{2M}{r}+\frac{8\pi P}{3}r^2\Bigl(1+\frac{3}{2}\frac{b}{r}\Bigr)\nonumber\\
&&-\frac{4\pi ab\log(r/b)}{r}-\frac{15\pi a}{2}\frac{b^2}{r^2}+O\bigl[(b/r)^3\bigr]\,.
\ea
Apart from the strange terms logarithmic and linear in $r$, we observe that the requirement for positivity of $a$,
signifying the attraction in between the molecules of the fluid, implies `spherical' horizon topology of the VdW black hole. Without loss of generality we can set $a=1/(2\pi)$.

\section{Conclusions}

\begin{figure}
\begin{center}
\rotatebox{-90}{
\includegraphics[width=0.34\textwidth,height=0.31\textheight]{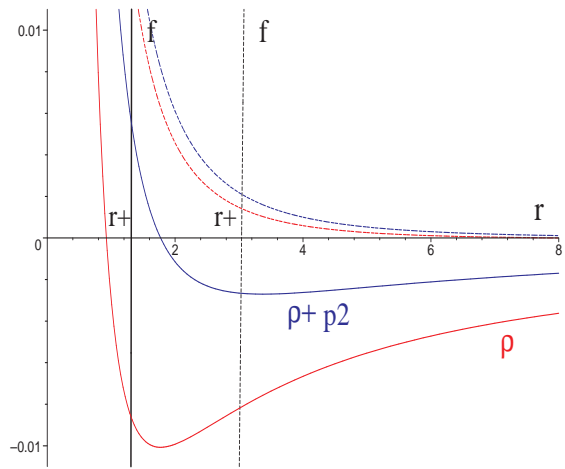}
}
\caption{{\bf Horizon and energy conditions.}
We display the energy density $\rho$ (red curves) and the quantity $\rho+p_2$ (blue curves); we always have $\rho+p_1=0$. The $x$-axis represents the radial coordinate $r$, the black curves depict metric function $f$, which is positive everywhere outside of $r_+$, giving rise to a black hole horizon. 
For large pressures $P$, none of the energy conditions are satisfied outside of the horizon, as illustrated by solid thick curves for $P=0.03$. As $P$ decreases energy conditions can be satisfied in the vicinity of the black hole horizon, see dotted curves for $P=0.001$.
In this figure, all dimensionful quantities are measured in units of the Van der Waals parameter $b$, while we have set  $M=0.1$ and $a=\frac{1}{2\pi}$. 
}  
\label{Fig1}
\end{center}
\end{figure}

The obtained metric \eqref{metric}  with $f$ given by \eqref{metf} reproduces exactly the Van der Waals equation \eqref{VdW} with the specific volume $v$, \eqref{v}, given by  
\be
v=2r_++3b\,. 
\ee
It is straightforward to show that the corresponding thermodynamic volume $V$ satisfies the reverse isoperimetric inequality \cite{CveticEtal:2010}. 
Unfortunately, the requirement for positivity of $b$  implies that none of the standard energy conditions are satisfied
everywhere outside of the horizon.  However it is possible to satisfy the
energy conditions within a region sufficiently close to the horizon provided the pressure is sufficiently small, as shown in Fig.~\ref{Fig1}.   
The effect of reducing the effective volume $v$ results in adding  `exotic matter' into the 
spacetime.\footnote{%
On the other hand, the increase of the effective volume can be achieved by adding matter that can obey all three standard energy conditions everywhere outside of the horizon. Consequently, it is easy to construct a black hole with thermodynamics mimicking the `anti-VdW equation' $T=(P+a/v^2)(v+b)$.}


We emphasize that our ansatz \eqref{ansatzh}
is  essentially unique.
It is straightforward to check that inclusion of higher powers of $P$ do not yield solutions consistent with the asymptotic  AdS structure postulated in \eqref{metric}.  Indeed our ansatz furnishes a
general procedure for constructing a metric from a given equation of state, at least in the spherically symmetric
case, and our approach can easily be extended to other dimensions. One can, for example, construct a black hole whose thermodynamics  coincides with the virial expansion of equation of state to an arbitrary order.

However our results also indicate that the asymptotic fall-off behaviour of the metric and matter fields will in general differ from that of standard electrovacuum and perfect fluid cases.   For the particular Van der Waals case we consider here,  an expansion of the stress-energy tensor at large $r$ yields
\beqa
\rho &=& -\frac{Pb}{r}+\frac{1-2\pi a}{8\pi r^2}+\frac{ab}{2r^3} + \cdots\,, \nonumber\\
p_2 &=& \frac{Pb}{2r} +\frac{ab}{4r^3} + \cdots\,,
\eeqa
whose  corresponding (exotic) matter content remains  to be found.   We also remark that, apart from the logarithmic term, 
the obtained metric is qualitatively similar to an exact solution to the field equations of a certain class of conformal
gravity theories \cite{Lu:2012xu} with vanishing stress-energy, affording an alternate interpretation of our
results.

Since the first law  \eqref{first}  is satisfied by \eqref{metric}  with $f$ given by \eqref{metf}, we can regard
$M$ as a ``mass''. An independent evaluation of this mass via either conformal \cite{Ashtekar:1984zz, Ashtekar:1999jx} or
counter term \cite{Mann:1999pc} methods is problematic due to   the weaker subleading falloff behaviour of the metric and the
stress-energy. It will be necessary to modify these approaches (somewhat along the lines of Dilaton gravity
\cite{Mann:2009id} or 
asymptotically Lifshitz metrics \cite{Ross:2009ar,Mann:2011hg}) to cancel the divergent terms that arise.

\subsection{Acknowledgments} 
We would like to thank T.~Delsate, P.~Pani, and M.R. Setare  for reading carefully our manuscript.
This research was supported in part by Perimeter Institute for Theoretical Physics and by the Natural Sciences and Engineering Research Council of Canada. Research at Perimeter Institute is supported by the Government of Canada through Industry Canada and by the Province of Ontario through the Ministry of Research and Innovation.


\begin{thebibliography}{10}

\bibitem{HawkingPage:1983}
S.~Hawking and D.~N. Page, {\it {Thermodynamics of Black Holes in anti-De
  Sitter Space}},  {\em Commun.Math.Phys.} {\bf 87} (1983) 577.

\bibitem{ChamblinEtal:1999a}
A.~Chamblin, R.~Emparan, C.~Johnson, and R.~Myers, {\it {Charged AdS black
  holes and catastrophic holography}},  {\em Phys.Rev.} {\bf D60} (1999)
  064018, [\href{http://xxx.lanl.gov/abs/hep-th/9902170}{{\tt
  hep-th/9902170}}].

\bibitem{Cvetic:1999ne}
M.~Cvetic and S.~Gubser, {\it {Phases of R charged black holes, spinning branes
  and strongly coupled gauge theories}},  {\em JHEP} {\bf 9904} (1999) 024,
  [\href{http://xxx.lanl.gov/abs/hep-th/9902195}{{\tt hep-th/9902195}}].

\bibitem{CaldarelliEtal:2000}
M.~M. Caldarelli, G.~Cognola, and D.~Klemm, {\it {Thermodynamics of
  Kerr-Newman-AdS black holes and conformal field theories}},  {\em
  Class.Quant.Grav.} {\bf 17} (2000) 399--420,
  [\href{http://xxx.lanl.gov/abs/hep-th/9908022}{{\tt hep-th/9908022}}].

\bibitem{Niu:2011tb}
C.~Niu, Y.~Tian, and X.-N. Wu, {\it {Critical Phenomena and Thermodynamic
  Geometry of RN-AdS Black Holes}},  {\em Phys.Rev.} {\bf D85} (2012) 024017,
  [\href{http://xxx.lanl.gov/abs/1104.3066}{{\tt arXiv:1104.3066}}].

\bibitem{KubiznakMann:2012}
D.~Kubiznak and R.~B. Mann, {\it {P-V criticality of charged AdS black holes}},
   {\em JHEP} {\bf 1207} (2012) 033,
  [\href{http://xxx.lanl.gov/abs/1205.0559}{{\tt arXiv:1205.0559}}].

\bibitem{Gunasekaran:2012dq}
S.~Gunasekaran, R.~B. Mann, and D.~Kubiznak, {\it {Extended phase space
  thermodynamics for charged and rotating black holes and Born-Infeld vacuum
  polarization}},  {\em JHEP} {\bf 1211} (2012) 110,
  [\href{http://xxx.lanl.gov/abs/1208.6251}{{\tt arXiv:1208.6251}}].

\bibitem{KastorEtal:2009}
D.~Kastor, S.~Ray, and J.~Traschen, {\it {Enthalpy and the Mechanics of AdS
  Black Holes}},  {\em Class.Quant.Grav.} {\bf 26} (2009) 195011,
  [\href{http://xxx.lanl.gov/abs/0904.2765}{{\tt arXiv:0904.2765}}].

\bibitem{Dolan:2010}
B.~Dolan, {\it {The cosmological constant and the black hole equation of
  state}},  {\em Class.Quant.Grav.} {\bf 28} (2011) 125020,
  [\href{http://xxx.lanl.gov/abs/1008.5023}{{\tt arXiv:1008.5023}}].

\bibitem{Altamirano:2014tva}
N.~Altamirano, D.~Kubiznak, R.~B. Mann, and Z.~Sherkatghanad, {\it
  {Thermodynamics of rotating black holes and black rings: phase transitions
  and thermodynamic volume}},  {\em Galaxies} {\bf 2} (2014) 89--159,
  [\href{http://xxx.lanl.gov/abs/1401.2586}{{\tt arXiv:1401.2586}}].

\bibitem{poisson2004relativist}
E.~Poisson, {\em A Relativist's Toolkit}.
\newblock Cambridge University Press, 2004.

\bibitem{CveticEtal:2010}
M.~Cvetic, G.~Gibbons, D.~Kubiznak, and C.~Pope, {\it {Black Hole Enthalpy and
  an Entropy Inequality for the Thermodynamic Volume}},  {\em Phys.Rev.} {\bf
  D84} (2011) 024037, [\href{http://xxx.lanl.gov/abs/1012.2888}{{\tt
  arXiv:1012.2888}}].

\bibitem{Lu:2012xu}
H.~Lu, Y.~Pang, C.~Pope, and J.~Vazquez-Poritz, {\it {AdS and Lifshitz Black
  Holes in Conformal and Einstein-Weyl Gravities}},  {\em Phys.Rev.} {\bf D86}
  (2012) 044011, [\href{http://xxx.lanl.gov/abs/1204.1062}{{\tt
  arXiv:1204.1062}}].

\bibitem{Ashtekar:1984zz}
A.~Ashtekar and A.~Magnon, {\it {Asymptotically anti-de Sitter space-times}},
  {\em Class.Quant.Grav.} {\bf 1} (1984) L39--L44.

\bibitem{Ashtekar:1999jx}
A.~Ashtekar and S.~Das, {\it {Asymptotically Anti-de Sitter space-times:
  Conserved quantities}},  {\em Class.Quant.Grav.} {\bf 17} (2000) L17--L30,
  [\href{http://xxx.lanl.gov/abs/hep-th/9911230}{{\tt hep-th/9911230}}].

\bibitem{Mann:1999pc}
R.~B. Mann, {\it {Misner string entropy}},  {\em Phys.Rev.} {\bf D60} (1999)
  104047, [\href{http://xxx.lanl.gov/abs/hep-th/9903229}{{\tt
  hep-th/9903229}}].

\bibitem{Mann:2009id}
R.~B. Mann and R.~McNees, {\it {Boundary Terms Unbound! Holographic
  Renormalization of Asymptotically Linear Dilaton Gravity}},  {\em
  Class.Quant.Grav.} {\bf 27} (2010) 065015,
  [\href{http://xxx.lanl.gov/abs/0905.3848}{{\tt arXiv:0905.3848}}].

\bibitem{Ross:2009ar}
S.~F. Ross and O.~Saremi, {\it {Holographic stress tensor for non-relativistic
  theories}},  {\em JHEP} {\bf 0909} (2009) 009,
  [\href{http://xxx.lanl.gov/abs/0907.1846}{{\tt arXiv:0907.1846}}].

\bibitem{Mann:2011hg}
R.~B. Mann and R.~McNees, {\it {Holographic Renormalization for Asymptotically
  Lifshitz Spacetimes}},  {\em JHEP} {\bf 1110} (2011) 129,
  [\href{http://xxx.lanl.gov/abs/1107.5792}{{\tt arXiv:1107.5792}}].

\end{thebibliography}

\newpage
\providecommand{\href}[2]{#2}\begingroup\raggedright\endgroup

\end{document}